\documentclass[aps,11pt,showpacs,nofootinbib]{revtex4}
\usepackage{tabularx} 
\usepackage{subfig} 
\usepackage{amsmath,amssymb,amsfonts,pifont,fancybox,float}
\usepackage{graphicx}
\usepackage[vcentermath]{youngtab}
\usepackage{epsfig}
\newcolumntype{C}{>{\displaystyle}c<{}}
\newcolumntype{L}{>{\displaystyle}l<{}}
\newcolumntype{R}{>{\displaystyle}r<{}}

\begin{document}
\title{Massless renormalization group flow in SU($N$)$_k$ perturbed conformal field theory}
\author{P. Lecheminant}
\affiliation{Laboratoire de Physique Th\'eorique et
Mod\'elisation, CNRS UMR 8089,
Universit\'e de Cergy-Pontoise, site de Saint-Martin, F-95300 Cergy-Pontoise Cedex, France.}
\begin{abstract}
We investigate the infrared properties of SU($N$)$_k$ conformal field theory perturbed by its
adjoint primary field in 1+1 dimensions.  The latter field theory is shown to govern the low-energy
properties of various SU($N$) spin chain problems. In particular, using a mapping onto $k$-leg  SU($N$) spin ladder, a massless renormalization group flow to SU($N$)$_1$ criticality  is predicted when $N$ and $k$ have no common divisor. The latter result extends the well-known massless flow between SU(2)$_k$ and SU(2)$_1$ Wess-Zumino-Novikov-Witten theories when $k$ is odd in connection to the Haldane's conjecture on SU(2)  Heisenberg spin chains.  A direct approach is presented in the simplest $N=3$ and $k=2$ case to investigate the existence of this massless flow.
\end{abstract}
\date\today
\maketitle     
\section{Introduction}
Conformal field theory (CFT) has attracted considerable interest over the years in problems ranging from
high-energy physics to statistical and condensed matter physics \cite{dms,mussardo,bookboso}.  In particular, 
it provides a full understanding of the physical properties of the emerging quantum criticality of one-dimensional (1D)
quantum problems. The low-energy relativistic spectrum of a 1D lattice model with a continuous
symmetry is described in terms of representations of a certain current algebra \cite{affleck85,affleck86}.
This affine symmetry determines the operator content of the theory
and all possible scaling dimensions of the operators are then
fixed by the conformal invariance of the underlying 
Wess-Zumino-Novikov-Witten (WZNW) model \cite{knizhnik,zamoloWZW}.
The CFT approach allows the computation of 
correlation functions, the finite-size spectrum as well as the entanglement
spectrum of the underlying lattice model \cite{cardy1,lauchli}.
The different critical phases are identified by
the central charge $c$ of the 
WZNW CFT which fixes the low-temperature behavior of the specific heat and the scaling of the entanglement entropy \cite{cardy,affleck86c,calabrese}.

On top of  the full description of the properties of the fixed point, the CFT approach
also gives access to the natural basis to investigate the effect of perturbations around it
\cite{zamoloint}.
As a result of a strongly relevant perturbation, the conformal symmetry might be lost and a mass gap 
is generated by the interaction. A second possible scenario is a massless renormalization group (RG) flow
where in the far infrared (IR) limit the conformal symmetry is restored with a smaller central charge
 \cite{zamolocth}.
 A well-known example of the latter phenomenon is the massless RG flow between 
 consecutive Virasoro minimal models ${\cal  M}_p$ 
and ${\cal  M}_{p-1}$  perturbed by a negative $\Phi_{13}$ relevant perturbation \cite{zamoloRG,ludwig,zamoloal,klassen}. 
 These RG flows might be studied by means of the power of integrability
 methods in case of integrable perturbations \cite{zamoloint,mussardo}.
 In absence of integrability, numerical approaches, as the truncated conformal space approach \cite{zamolotruncated} or 
 his improvement \cite{konik} are efficient methods to fully determine the IR properties of a perturbed
 CFT.
 
 In this paper, we investigate the possible occurrence of a massless RG flow for 
 the SU($N$)$_k$ WZNW CFT with central charge $c= k (N^2 -1)/(N+k)$ perturbed by its adjoint primary field 
 $\Phi_{\rm adj}$ with scaling dimension $2N/(N+k)$. The Hamiltonian density of the resulting model is defined 
 as follows:
 \begin{equation}
{\cal H} = \frac{2\pi}{N+k}\Big(:I_R^A I_R^A: + :I_L^A I^A_L:\Big) + \lambda \;  {\rm Tr}  \; \Phi_{\rm adj} ,
\label{model}
\end{equation}
where $I_{R,L}^A, A= 1, \ldots, N^2 -1$ are the chiral currents which satisfy the SU($N$)$_k$ current algebra:
\begin{eqnarray}
I_{L}^A\left(z\right) I_{L}^B\left(\omega\right) &\sim& \frac{k \delta^{AB}  }{8 \pi^2 
\left(z - \omega \right)^2} + 
\frac{i f^{ABC} }{2 \pi \left(z - \omega \right)} I_{L}^C\left(\omega\right),
\label{curralgSUNk}
\end{eqnarray}
with a similar definition for the right current and $f^{ABC}$ are the structure constants of the SU($N$) group.
In Eq. (\ref{model}), $:O:$ denotes the normal ordering of operator $O$ and a summation over repeated
indices is assumed throughout the paper. 
The adjoint primary field can be expressed in terms of the SU($N$)$_k$ WZNW field $G$
with scaling dimension $(N^2 -1)/N(N+k)$  \cite{knizhnik}:
\begin{equation}
\Phi^{AB}_{\rm adj} \sim {\rm Tr} ( G^{\dagger} T^{A} G T^{B} ), 
\label{adjointfield}
\end{equation}
$T^{A}$ being the SU($N$) generators transforming in the fundamental representation normalized according
to $ {\rm Tr} (T^{A} T^{B}) =  \delta^{AB}/2$.

The main physical motivation to study the IR properties of the Hamiltonian (\ref{model}) 
stems from its application to 1D Heisenberg spin chain models.
When $N=2$, it is well-known that model  (\ref{model}) with $k=2S$ accounts for the IR properties of 
spin-$S$ Heisenberg chain  \cite{affleckhaldane}.
A massless flow with emerging SU(2)$_1$ quantum criticality is obtained for $\lambda >0$ and
odd $k$ whereas a massive behavior occurs in other cases \cite{affleckhaldane,tsvelik}.
This result is directly related to the famous Haldane's conjecture that integer Heisenberg spin chain has a spectral gap
while in the half-integer case a massless behavior in the SU(2)$_1$ universality class is stabilized \cite{haldane}.
For general $N$, it has been recently proposed that model  (\ref{model}) with $k=2$ governs the 
quantum phase transition between dimerized and  1D SU($N$)  symmetry-protected topological phases  \cite{Nonne2013,furusaki,bois}.

In this paper,  it will be shown that the SU($N$)$_k$  perturbed CFT (\ref{model}) describes
the low-energy limit of weakly coupled SU($N$)  spin ladder and SU($N$) spin chain models with symmetric
rank-$k$ tensor representation. We will investigate the IR properties of model (\ref{model}) and try to extend the known $N=2$ results to general $N$. We will see that the field theory (\ref{model}) is massive for all $N$ and $k$ when $\lambda <0$. In contrast, when $\lambda > 0$, using a mapping onto $k$-leg SU($N$) spin ladder, a massless RG flow to SU($N$)$_1$ CFT is expected when $N$ and $k$ have no common divisor. In the simplest
$k=2$ and $N=3$  case we perform a direct approach by means of Gepner's parafermions (GP) \cite{gepnerpara}.
In this respect, we conclude that the SU(3)$_2$ CFT perturbed by its adjoint primary field enjoys a massless
RG flow down to the SU(3)$_1$ universality class.

The rest of the paper is organized as follows. In Sec. II, we present the mapping of model (\ref{model})
onto  $k$-leg SU($N$) spin ladder. Using known results in SU($N$) spin chains, we deduce our main
conclusion on the IR properties of the field theory (\ref{model}). In Sec. III, a direct approach is developed
in the simplest $k=2$ and $N=3$ case. Finally, our concluding remarks are given in Sec. IV and the paper is supplied with 
two appendices which provide additional information.

\section{Mapping onto $k$-leg SU($N$) spin ladder}

In this section, we show that model (\ref{model}) describes the low-energy physics of weakly coupled 
 $k$-leg SU($N$) spin ladder with lattice Hamiltonian:
 \begin{equation}
{\cal H}_{\rm ladder} =  J_{\parallel} \sum_i  \sum_{l=1}^{k} S^{A}_{l,i}  S^{A}_{l,i+1} 
+   J_{\perp} \sum_i  \sum_{l=1}^{k-1} S^{A}_{l,i}  S^{A}_{l+1,i}  ,
\label{2leg}
\end{equation}
 where $S^{A}_{l,i}$ denote the SU($N$) spin operators, which transform in the fundamental
 representation of SU($N$) (represented by the Young diagram {\tiny $ \yng(1)$}), 
 on the ith site  and the $l=1,\ldots, k$ leg of the spin ladder. 
 We assume that the ladder has open transverse boundary conditions, i.e., the outer spins $S^{A}_{N,i}$ 
 do not interact with $S^{A}_{1,i}$ ones (see Fig. 1). The spin ladder (\ref{2leg}) can, in principle, be manufactured 
 in the context of ultracold alkaline-earth or ytterbium atoms where the existence of an high
 SU($N$) symmetry has been recently demonstrated experimentally \cite{gorshkov,Cazalilla-H-U-09,bloch,rey,Cazalilla-R-14,capponirevue}.
 In absence of the interchain coupling ($J_{\perp} = 0$), the model is $k$ decoupled Sutherland models \cite{sutherland}. 
 The latter is integrable and displays a quantum critical behavior in the SU($N$)$_1$ universality
 class with central charge $c= N-1$ \cite{affleck86,affleck}.
 In the weak-coupling limit $ |J_{\perp}| \ll J_{\parallel}$, one can then perform a low-energy approach
 to deduce the physical properties of the spin ladder (\ref{2leg}). In this respect, we will extend
 the results of Ref. \onlinecite{phletsvelik} to the general $k >2$ case.
\begin{figure}[ht]
\begin{center}
\includegraphics[width=0.8\columnwidth,clip]{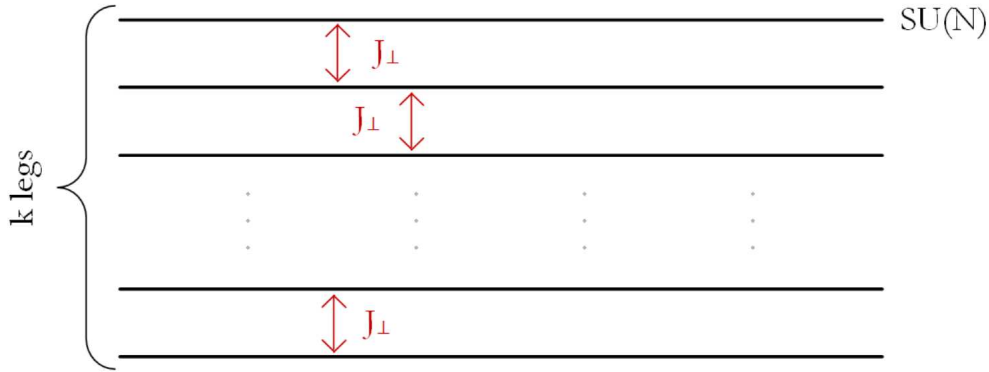}
\caption{(color online) $k$-leg SU($N$) spin ladder.}
\label{fig:Nleg}
\end{center}
\end{figure}

\subsection{Continuum limit}

 The low-energy properties of the Sutherland model can be obtained by starting from
the U($N$) Hubbard model at $1/N$ filling with large repulsive $U$ interaction \cite{affleck,assaraf,itoi,manmana}.
At low-energy below the charge gap, the SU($N$) operators in the continuum limit are
described by \cite{affleck,assaraf,itoi}:
\begin{equation}
S^{A}_{l,i} \simeq J^{A}_{l L} +  J^{A}_{l R} + \mbox{e}^{ i 2k_F x} N^{A}_l + \mbox{e}^{-i 2k_F x} N^{A \dagger}_l 
+  \mbox{e}^{ i 4k_F x} n^{A}_l  + ..,
\label{spinop}
\end{equation}
where $x= ia_0$ ($a_0$ being the lattice spacing) and the Fermi momentum is $k_F = \pi/Na_0$
since the underlying Hubbard model is $1/N$ filled (one fermion per site).
In Eq. (\ref{spinop}), $J^{A}_{l L,R}$ are the left and right SU($N$)$_1$ currents,
the $2k_F$ and $4k_F$ parts are related to the SU($N$)$_1$ primary fields which
transform respectively in the fundamental representation ({\tiny $ \yng(1)$}) and the antisymmetric one  ({\tiny $ \yng(1,1)$})
with dimension $N(N-1)/2$  of 
the SU($N$) group. The SU($N$)$_1$ currents can be expressed in terms of the underlying  left (right)-moving Dirac fermions
 \cite{bookboso}:
\begin{equation}
J^A_{l R} = R_{l \alpha}^{\dagger} T^A_{\alpha \beta}
R_{l \beta} , \; \;  
J^A_{l L} = L_{l \alpha}^{\dagger} T^A_{\alpha \beta}
L_{l \beta} ,
\label{suNcur}
\end{equation}
where, here, there is no sum on $l$ and we have $\alpha,\beta = 1, \ldots, N$.
The $2k_F$ term of Eq. (\ref{spinop}) reads as follows in terms of the Dirac fermions:
\begin{eqnarray}
N^{A}_l = \langle L_{l \alpha}^{\dagger}  T^A_{\alpha \beta} R_{l \beta}  \rangle_c = 
 \langle \mbox{e}^{i\sqrt{4 \pi/N} \Phi_{lc}} \rangle_c \; {\rm Tr} ( g_l  T^A) = \lambda   \; {\rm Tr} ( g_l  T^A),
 \label{2kf}
\end{eqnarray}
where $\langle A \rangle_c$ denotes an average over the charge degrees of freedom which are fully gapped
in the large $U$ limit, and $ \lambda   = \langle  \mbox{e}^{i\sqrt{4 \pi/N} \Phi_{lc}} \rangle_c \ne 0$ can be chosen real for a matter of convenience.
The field $g_l$ is the SU($N$)$_1$ WZNW primary field which transforms
in the fundamental representation of SU($N$). 
In the non-Abelian bosonization approach \cite{knizhnik,witten,affleck86},
it is described (see Eq. (\ref{2kf})) by the identity:
\begin{eqnarray}
 g_{l \beta \alpha} \sim \langle   \mbox{e}^{-i\sqrt{4 \pi/N} \Phi_{lc}} 
 L_{l \alpha}^{\dagger}  R_{l \beta}  \rangle_c ,
 \label{grep}
\end{eqnarray}
and has scaling dimension $(N - 1)/N$.

With these basic facts at hands, the continuum description of the decoupled SU($N$) spin ladder
is given by the Hamiltonian density:
\begin{equation}
{\cal H}_0 =  \frac{2\pi v}{N + 1} \left[ : J^A_{l R} J^A_{l R}: + : J^A_{l L} J^A_{l L}: 
\right] - \gamma J^A_{l R} J^A_{l L} ,
\label{contfreeham}
\end{equation}
where $v$ is the spin velocity and
$\gamma > 0$ so that the perturbation is  a marginal irrelevant current-current term.
We will discard this perturbation in the following which gives logarithmic corrections \cite{affleckschulz}.
In the weak-coupling limit, one can use the low-energy description
(\ref{spinop}) to get the continuum limit of the $k$-leg SU($N$) spin ladder.
The leading contribution stems from the $2k_F$ part of the spin operator (\ref{spinop}) and we find:
\begin{equation}
{\cal H}_{\rm ladder} =    \frac{2\pi v}{N + 1} \left[ : J^A_{l R} J^A_{l R}: + : J^A_{l L} J^A_{l L}: 
\right] 
+ J_{\perp} \lambda^2 \sum_{l=1}^{k-1} \left({\rm Tr}( g_l  T^A) \; {\rm Tr}( g^{\dagger}_{l+1}  T^A) + H.c.\right) ,
\label{contham}
\end{equation}
where the perturbation is strongly  relevant with scaling dimension $2 (N-1)/N < 2$.
The leading interaction in the weak-coupling regime takes thus the form of $k$ coupled SU($N$)$_1$ WZNW models. 
Using the SU($N$) identity:
\begin{equation}
 \sum_A T^A_{\alpha \beta} T^A_{\gamma \rho} = 
\frac{1}{2} \left(\delta_{\alpha \rho} \delta_{\beta \gamma}
- \frac{1}{N}\; \delta_{\alpha \beta} \delta_{\gamma \rho} \right),
\label{SUNident}
\end{equation}
it is then useful to rewrite the interacting part of model (\ref{contham}) in the following form:
\begin{eqnarray}
{\cal H}_{\rm int} &=& {\cal H}_1 + {\cal H}_2 \label{Vint} \\
{\cal H}_1 &=& \lambda_1 \sum_{l=1}^{k-1} \left( {\rm Tr} ( g_l  g^{\dagger}_{l+1}) + H.c.\right) \label{V1}  \\
{\cal H}_2 &=& \lambda_2 \sum_{l=1}^{k-1} \left( {\rm Tr} ( g_l) {\rm Tr}(g^{\dagger}_{l+1}) + H.c.\right) , 
\label{V12}
\end{eqnarray}
with $\lambda_2 = - \lambda_1/N$ and $\lambda_1 = J_{\perp} \lambda^2 /2$.

\subsection{Strong-coupling arguments}
 
The two strongly relevant perturbations in Eq. (\ref{Vint}) are of very different nature.
Indeed, one observes that ${\cal H}_1$ is invariant under an SU($N$)$_{\rm L}$ $\times$ 
SU($N$)$_{\rm R}$ symmetry: $g_l \rightarrow U_{\rm L} g_l  U_{\rm R}$, $U_{\rm L,R}$ being two independent
SU($N$) matrices. In stark contrast, ${\cal H}_2$ is only SU($N$) invariant: $g_l \rightarrow U g_l  U^{\dagger}$,
with $U$ belonging to SU($N$). In close parallel to the $N=2$ case \cite{phle,Nonne2012}, one way to separate the different degrees of freedom of the problem is to consider the following CFT embedding, built from the product of $k$ SU($N$)$_1$  CFTs:
\begin{equation}
SU(N)_1 \times SU(N)_1 \ldots \times SU(N)_1 \sim SU(N)_k \times  \mathbb{G}_{N,k} ,
\label{embedding}
\end{equation}
where $ \mathbb{G}_{N,k}$ is a coset CFT with central charge $c= k(N-1)(k - 1)/(N+k)$. In the $k=2$ case, 
$ \mathbb{G}_{N,2}$ corresponds to the  $\mathbb{Z}_{N}$ parafermionic CFT \cite{para,gepner}.
For general $k$,  $\mathbb{G}_{N,k}$ is the sum of $k-1$ consecutive coset models CFTs 
$SU(N)_p \times SU(N)_1/SU(N)_{p+1}$ which enjoy an extended 
$\mathbb{W}_{N}$ symmetry \cite{wCFT}.
The SU($N$)$_k$ CFT is generated by SU($N$)$_k$ chiral currents $I_{R,L}^A$ which are
the sum of $k$ SU($N$)$_1$ currents: $I_{R,L}^A = \sum_{l=1}^{k} J^A_{l R,L}$.
Since ${\cal H}_1$ is SU($N$)$_{\rm L}$ $\times$ 
SU($N$)$_{\rm R}$ invariant, one expects that 
it  does not depend on the SU($N$)$_k$ CFT but only on the $ \mathbb{G}_{N,k}$ CFT.
In fact, one can check using the following operator product expansion (OPE) for the left current \cite{dms}:
\begin{eqnarray}
J_{l L}^A\left(z\right)  (g_{p})_{\alpha \beta} (0,0) &\sim& - \frac{\delta_{lp}}{2 \pi z} \; T^{A}_{\alpha \gamma}
(g_{p})_{\gamma \beta} (0,0) \nonumber \\
 J_{l L}^A\left(z\right)  (g^{\dagger}_{p})_{\beta \alpha} (0,0) &\sim& \frac{\delta_{lp}}{2 \pi z}  \;
 (g^{\dagger}_{p})_{\beta \gamma} (0,0)
 T^{A}_{\gamma \alpha},
\label{OPEWZW}
\end{eqnarray}
that $I_{L}^A\left(z\right) {\cal H}_1 \sim 0$, i.e., ${\cal H}_1$ cannot depend on 
the SU($N$)$_k$ primary fields. In contrast, ${\cal H}_1$  is a relevant primary field of the  $\mathbb{G}_{N,k}$ CFT
with scaling dimension $2(N-1)/N$ and is expected to open a gap $\Delta$ for these discrete degrees of freedom.
However, depending on the sign of the coupling constant $\lambda_1$, the relevant perturbation might
give a massless RG flow as the one between consecutive minimal models  
 \cite{zamoloRG,ludwig,zamoloal,klassen}.
When $k=2$, it has been shown in Ref. \onlinecite{phletsvelik} that ${\cal H}_1$ is in fact related to a perturbation 
of the $\mathbb{Z}_{N}$  CFT:
\begin{equation}
{\cal H}_1 = {\cal H}^{0}_{{\mathbb Z}_N}    + \lambda_1  \left( \Psi_{1L}  \Psi_{1R}  + H.c. \right),  
\label{paraintdeforfateev}
\end{equation}
where ${\cal H}^{0}_{{\mathbb Z}_N}$ is the Hamiltonian density of the  ${\mathbb{Z}}_N$ parafermionic CFT
which is generated by the chiral currents $ \Psi_{1L,R}$ with conformal weights $h, {\bar h} = (N-1)/N$ \cite{para}.
Model (\ref{paraintdeforfateev}) is known to be an integrable massive field theory for all sign of $\lambda_1$ when $N$ is even \cite{fateev}.
In contrast, when $N$ is odd, it displays a massless RG flow from the ${\mathbb{Z}}_N$ fixed point
to the minimal model series ${\cal M}_N$  in the IR limit when $\lambda_1>0$, 
while for $\lambda_1< 0$,  it is again a massive field theory \cite{fateev,fateevzamolo}.
In the following, we thus consider only the latter case, i.e., $J_{\perp} <0$, for general $N$ and $k$, where a spectral gap  $\Delta$ is expected to be formed for the $\mathbb{G}_{N,k}$ degrees of freedom.

In the low-energy limit $E \ll \Delta$, the IR properties of the spin ladder model (\ref{2leg}) are then governed
by the SU($N$)$_k$ CFT with a certain perturbation which stems from the ${\cal H}_2$ contribution in Eq. (\ref{Vint}).
To perform this low-energy limit, it is convenient to consider the full Euclidean action of 
model (\ref{Vint}). In this respect, we denote by $W(g)$ the action of the SU($N$)$_1$ WZNW model:
\begin{eqnarray}
W(g) &=& \frac{1}{8\pi} \int d^2 x \;  {\rm Tr}  (\partial^{\mu} g^{\dagger} \partial_{\mu} g) 
+ \Gamma(g) \nonumber \\
\Gamma(g) &=& \frac{-i}{12\pi}   \int_B d^3 y \; \epsilon^{\alpha \beta \gamma} 
 {\rm Tr} (g^{\dagger} \partial_{\alpha} g g^{\dagger} \partial_{\beta} g g^{\dagger} \partial_{\gamma} g),
\label{WZW}
\end{eqnarray}
where $g$ is an SU($N$) field and $\Gamma(g)$ is the famous WZNW topological term.
The 	action of the $k$-leg SU($N$)  spin ladder problem is then:
\begin{equation}
{\cal S} = \sum_{i=1}^{k} W(g_i)  +  \lambda_1 \sum_{l=1}^{k-1} \int d^2 x  \left( {\rm Tr} ( g_l  g^{\dagger}_{l+1}) + H.c.\right) 
+ \lambda_2 \sum_{l=1}^{k-1} \int d^2 x  \left( {\rm Tr} ( g_l) {\rm Tr}(g^{\dagger}_{l+1}) + H.c.\right) .
\label{actionkleg}
\end{equation}

For a ferromagnetic interchain coupling $J_{\perp} < 0$, we have
$\lambda_1 < 0$ and the second term in the right-hand side of Eq. (\ref{actionkleg}) opens a spectral gap  $\Delta$ as discussed above. In the strong-coupling regime, the configuration
$g_{l+1} =  g_l$ ($l = 1, \ldots, k -1$)  minimizes the term with coupling constant $\lambda_1$ in  Eq. (\ref{actionkleg}). 
One can then integrate out these degrees of freedom to get an 
effective action for the $g_1$ field when $E \ll \Delta$:
\begin{equation}
{\cal S}_{\rm eff} =  k W(g_1) + \lambda   \int d^2 x \; : {\rm Tr} g_1  \; {\rm Tr} g_1^{\dagger} :,
\label{actiongeff}
\end{equation}
with $ \lambda = k \lambda_2 > 0$. One observes that the effective action (\ref{actiongeff})
describes an SU($N$)$_k$ WZNW model perturbed by the ${\rm Tr} g_1  \; {\rm Tr} g_1^{\dagger}$ field
with a positive coupling constant.
The latter perturbation corresponds to the trace of the adjoint field $ {\rm Tr}  \; \Phi_{\rm adj} $ as it can be seen
from the definition (\ref{adjointfield}) and the identity (\ref{SUNident}).

\subsection{Conjecture}

The low-energy properties of the $k$-leg SU($N$) spin ladder with a ferromagnetic interchain coupling $J_{\perp} <0$
are thus captured by the perturbed  SU($N$)$_k$ WZNW model (\ref{model}) or equivalently
by the action (\ref{actiongeff}). When $\lambda <0$, one can immediately argue from Eq. (\ref{actiongeff})  that in the
ground state: $g_1 =  e^{ i 2 \pi k /N} I$, $k =1, \dots, N$, i.e., elements of the center of the  SU($N$) group.  Fluctuations around these minima give rise to massive degrees of freedom. When $\lambda > 0$,  the situation is less clear and 
a massless RG flow is an intriguing possibility. In this respect, the connection to $k$-leg SU($N$)  spin ladder might give
some interesting information. In the $N=2$ case, it is well established that an adiabatic continuity occurs between weak and strong coupling limits for all sign of $J_{\perp}$  \cite{bookboso,Giamarchi-book}.
The $k$-leg SU(2) spin ladder with ferromagnetic interchain coupling $J_{\perp} < 0$ is known to be gapless (respectively
fully gapped) with one gapless bosonic mode, i.e., $c=1$  when $k$ is odd (respectively even) \cite{Giamarchi-book}.
The latter result is in full agreement with the massless RG flow of model (\ref{actiongeff}) from SU(2)$_k$ to SU(2)$_1$
when  $k$ is odd \cite{affleckhaldane,tsvelik}.

Unfortunately, there are no known numerical results for SU($N$) spin ladder  with $N>2$ and general $k$ 
except for a two-leg SU(3) spin ladder where an adiabatic continuity has been shown numerically when $J_{\perp} <0$
 \cite{capponiSU3}. Assuming such continuity when  $J_{\perp} < 0$ for all $N$, the nature of the IR properties of model  (\ref{actiongeff}) when $N>2$ can then be inferred from a lattice strong-coupling limit of model (\ref{2leg}) with $J_{\perp} \rightarrow - \infty$.
In the latter limit, the $k$-leg SU($N$) spin ladder (\ref{2leg}) is equivalent 
to a single 1D SU($N$) spin chain where the spin operator belongs to the $k$th symmetric representation  of SU($N$) which is described by a Young tableau with one line of $k$ boxes.
The phase diagram of 1D SU($N$)  Heisenberg spin chain in different representations is rather well understood  
\cite{affleck,lieb,sachdev,greiter,nonne2011,dufour}.
If $k$ and $N$ have no common divisor, the SU($N$) Heisenberg spin chain in the $k$th 
symmetric representation is known to display an SU($N$)$_1$ quantum critical behavior \cite{affleck,greiter}.
When $N=k$ a spin gap phase is expected while if $k$ and $N$ have a common divisor different from $N,$
the situation is less clear.
We will thus consider here only $k$-leg SU($N$) spin ladder when $k$ and $N$ have no common divisor.

This result, together with the identification (\ref{actiongeff}), leads us to the conjecture that the 
 SU($N$)$_k$ WZNW model (\ref{model}) perturbed by its adjoint primary field has a massless RG flow for $\lambda >0$
 to SU($N$)$_1$ CFT  if $k$ and $N$ have no common divisor.

 It might be interesting to relate the proposed massless RG flow to symmetry protection 
 of critical phases with an SU($N$) symmetry. Recently, the massless RG flows of SU(2)$_k$ WZNW models
 have been classified non-perturbatively thanks to a selection rule based on the global anomaly of 
 the ${\mathbb{Z}}_2$  discrete symmetry of the center of the SU(2) group \cite{oshikawa}.
 For perturbations preserving SU(2) and ${\mathbb{Z}}_2$ symmetries, a massless RG flow
 between SU($2$)$_k$ and SU($2$)$_{k'}$ can occur if only if $k$ and $k'$ have the same parity due to
 anomaly matching mecanism  \cite{oshikawa}.
 The latter result stems from the fact that the ${\mathbb{Z}}_2$ orbifold of the SU(2)$_k$ WZNW model is a consistent
 CFT, i.e., modular invariant, without global anomaly only if $k$ is even \cite{gepnerwitten}.
 Anomaly matching requires then a selection rule for the massless RG flow between SU(2)$_k$ WZNW 
 models \cite{oshikawa}.
 For perturbations invariant under SU(2) and ${\mathbb{Z}}_2$ symmetries, a massless RG flow
 between  SU(2)$_k$ and SU(2)$_1$ WZNW models is only possible when $k$ is odd  \cite{oshikawa}.
 It might be interesting to extend this argument for a perturbation invariant under SU($N$). On top of the latter symmetry,
 the field theory (\ref{model}) enjoys an ${\mathbb{Z}}_N$ symmetry, $G \rightarrow   e^{ i 2 \pi /N}  G$, 
 which corresponds to the one-step translation symmetry of the underlying SU($N$) spin ladder.
 In this respect, in close parallel to the $N=2$ case, described in Ref. \onlinecite{oshikawa}, 
 we consider an ${\mathbb{Z}}_N$ orbifold of SU($N$)$_k$ WZNW models. 
 The spectrum of these models have been determined in Ref. \onlinecite{felder}.
 When $N$ is odd  modular invariants exist for all $k$ while for even $N$, $k$ should be even to define a
 consistent CFT \cite{felder}.
 The latter result gives a selection rule for the massless RG flow between SU($N$)$_k$ WZNW theories 
 with ${\mathbb{Z}}_N$ invariant perturbation as the field theory (\ref{model}).
 In particular, we observe that the conjectured massless flow  SU($N$)$_k$ $\rightarrow$ SU($N$)$_1$
 when $k$ and $N$ have no common divisor is compatible with the anomaly matching mechanism of
 the ${\mathbb{Z}}_N$ symmetry.
 
 It is worth noting that the IR massless flow for $k=2$ might be explored 
 perturbatively in the large-$N$ limit. The scaling dimension of the SU($N$)$_2$ perturbed adjoint field is indeed very close
to two when $N \gg 1$: $\epsilon = 2 - \Delta_{\rm adj}  = 4/(N+2) \ll 1$. 
A perturbative RG approach is thus called for to find a non-trivial fixed point for model (\ref{model}) with $k=2$ in
the large-$N$ limit. 
Such analysis is similar in spirit
to the  one-loop RG approach of the massless flow between consecutive minimal models ${\cal M}_p$  and
${\cal M}_{p-1}$  induced by the $\Phi_{13}$ perturbation when $p \rightarrow \infty$ \cite{zamoloRG,ludwig}.
The analogy can be made more precise by considering the SU($N$)$_2$  fusion rules of $\Phi_{\rm adj}$ with itself
which occurs at the one-loop level.
The latter  can be derived by exploiting  the level-rank duality and the fusion rules of SU(2)$_N$ \cite{dms,schnitzer}:
\begin{equation}
\Phi_{\rm adj} \times \Phi_{\rm adj} \sim I + \Phi_{\rm adj} + \Phi^{'},
\label{fusionrule}
\end{equation}
where $\Phi^{'}$ corresponds to an SU($N$)$_2$ primary field when $N>3$ which transforms in a representation of SU($N$) described by the following Young tableau:
\begin{equation}
\text{\scriptsize $N-2$} \left\{ 
\yng(2,2,1,1)
 \right.   \; .
 \end{equation}
The primary field $\Phi^{'}$ has scaling dimension $\Delta = 4(N-1)/(N+2) >2$ and is 
an irrelevant contribution when $N >4$.
The situation is thus in striking parallel to the minimal model perturbed by the $\Phi_{13}$ primary field which
enjoys the fusion rule: $\Phi_{13} \times \Phi_{13} \sim I + \Phi_{13} + \Phi_{15}$, 
$ \Phi_{15}$ being an irrelevant field \cite{dms}.
We then expect a non-trivial fixed point in the large-$N$ limit. The details of the perturbative
analysis will be presented elsewhere \cite{raoul}.

Finally, in Appendix A, we relate the field theory  (\ref{model}) with $k=2$ to a single SU($N$) spin chain problem where the spin operators belong to symmetric representation ({\tiny $\yng(2)$}) with dimension $N(N+1)/2$
of the  SU($N$) group. It paves the way to the direct numerical investigation of the conjecture for finite $N$ in the simplest $k=2$ case.

\section{Parafermionic approach when $k=2$ and $N=3$}

In this section, we investigate the conjectured massless RG flow for model (\ref{model}) by means of an direct approach in the simplest case, i.e., $k=2$ and $N=3$.
The approach is based on the Gepner's parafermions \cite{gepnerpara} 
which extends the  ${\mathbb{Z}}_k$ parafermionic approach \cite{cabra} of the known massless RG flow for $N=2$, which is reviewed in Appendix B. In this respect, it is  useful  to consider the following conformal embedding:
\begin{equation}
SU(3)_2 \sim \frac{SU(3)_2}{U(1)^{2}} \times U(1)^{2} ,
\label{gepneremb}
\end{equation}
where the coset SU(3)$_2/ U(1)^{2}$ describes the so-called GP CFT with central
charge $c= 6/5$ \cite{gepnerpara}.
The SU(3)$_2$ primary field $G^{{\vec \Lambda}, {\vec {\bar \Lambda}}}_{{\vec \lambda}, {\vec {\bar \lambda}}}$ 
transforms  in the
left and right SU($3$) representations with highest weights ${\vec \Lambda}$ and ${\vec {\bar \Lambda}}$, ${\vec \lambda}$ and ${\vec {\bar \lambda}}$ being weights respectively in the ${\vec \Lambda}$ and ${\vec {\bar \Lambda}}$ representations. Introducing two left and right bosonic fields $\vec \Phi_{L,R} = (\Phi_{1L,R}, \Phi_{2L,R})$, one can relate these primary fields
to the one in the GP CFT \cite{gepnerpara}:
\begin{equation}
 G^{{\vec \Lambda}, {\vec {\bar \Lambda}}}_{{\vec \lambda}, {\vec {\bar \lambda}}} \sim : \exp \left(
 i \sqrt{2 \pi} \; {\vec \lambda} \cdot {\vec \Phi_{L}}
 + i \sqrt{2 \pi}  \;  {\vec {\bar \lambda}} \cdot {\vec \Phi_{R}} \right):
  \Phi^{{\vec \Lambda}, {\vec {\bar \Lambda}}}_{{\vec \lambda}, {\vec {\bar \lambda}}}
 ,
\label{gepnerprimary}
\end{equation}
where $ \Phi^{{\vec \Lambda}, {\vec {\bar \Lambda}}}_{{\vec \lambda}, {\vec {\bar \lambda}}}$ 
denotes the GP primary field with holomorphic dimension:
\begin{equation}
h^{{\vec \Lambda}}_{{\vec \lambda}} = \frac{{\vec \Lambda} \cdot ({\vec \Lambda} + 2 \vec \rho)}{10} - \frac{{\vec \lambda} \cdot {\vec \lambda}}{4} ,
\label{dimGP}
\end{equation}
$2 \vec \rho$ being the sum of all positive roots of the Lie algebra of SU(3): su(3).  
The SU(3)$_2/ U(1)^{2}$ GP primary fields have the following identification \cite{gepnerident}:
\begin{eqnarray}
 \Phi^{\Lambda_1, \Lambda_2}_{\lambda_1, \lambda_2} &=& \Phi^{2 - \Lambda_1
- \Lambda_2, \Lambda_1}_{\lambda_1+ 2, \lambda_2} =   \Phi^{ \Lambda_2, 2 - \Lambda_1
- \Lambda_2}_{\lambda_1, \lambda_2+2} \nonumber \\
\Phi^{\Lambda_1, \Lambda_2}_{\lambda_1, \lambda_2} &=& \Phi^{\Lambda_1, \Lambda_2}_{\lambda_1+ 4, \lambda_2 -2}
=  \Phi^{\Lambda_1, \Lambda_2}_{\lambda_1 - 2, \lambda_2 +4} ,
\label{GPSU3ident}
\end{eqnarray}
where $\lambda_{1,2}, \Lambda_{1,2},$ are Dynkin labels and for notational clarity, we have omitted the 
weights in the right sector  for the GP and SU(3)$_2$ primary fields. 
The identification (\ref{GPSU3ident}) leads to eight GP primary fields \cite{ardonne}:
 $\{I,\psi_{1}, \psi_{2}, \psi_{3},   \sigma_1,  \sigma_2,  \sigma_{3}, \rho \}$.
The GP primary fields, which appear in the expression of the 
adjoint SU(3)$_2$ primary field $G^{1,1}$,  are together with their holomorphic dimensions:
\begin{eqnarray}
\rho &=&  \Phi^{1, 1}_{0, 0}, \; \; h_{\rho} = 3/5 \nonumber \\
\sigma_1 &=& \Phi^{1, 1}_{-1, 2}, \; \; h_{\sigma_1} = 1/10 \nonumber \\
\sigma_2 &=& \Phi^{1, 1}_{2, -1}, \; \; h_{\sigma_2} = 1/10 \nonumber \\
\sigma_3 &=& \Phi^{1, 1}_{1, 1}, \; \; h_{\sigma_3} = 1/10 .
\label{GPSU3}
\end{eqnarray}
These fields are Hermitean operators due to the identification (\ref{GPSU3ident}).
The fusion rules between these primary fields have been derived in Ref. \onlinecite{ardonne}:
\begin{eqnarray}
\rho \times \rho &=&  I +  \rho,  \;  \sigma_i   \times \sigma_i  = I +  \rho, \; 
\sigma_1   \times \sigma_2  = \psi_{3} + \sigma_3, \; \sigma_1   \times \sigma_3  = \psi_{1} + \sigma_2, \;
 \nonumber \\
 \sigma_2   \times \sigma_3  &=& \psi_{2} + \sigma_1, 
\rho \times \sigma_1    = \psi_{2} + \sigma_1, \;  \rho \times \sigma_2   = \psi_{1} + \sigma_2, \;
\rho \times \sigma_3    = \psi_{3} + \sigma_3, 
\label{GPSU3fusion}
\end{eqnarray}
where the $\psi_i$ are three GP primary fields with holomorphic dimension $1/2$ which can be chosen as:
$\psi_1 = \Phi^{0, 0}_{2, -1}, \psi_2 = \Phi^{0, 0}_{1, -2}$ and $\psi_3 = \Phi^{0, 0}_{1, 1}$.

With all these results, one can derive some representations of the  SU(3)$_2$ primary fields
in terms of two bosons and the GP primary fields. Let us first consider the fundamental representation of SU(3) with highest weight $\vec \Lambda = (1,0)$ and weights ${\vec \lambda}$: $\{(1,0), (-1,1), (0,-1)\}$.
Using the decomposition (\ref{gepnerprimary}), we find the expression of the trace of the SU(3)$_2$ WZNW primary $G$
with scaling dimension $8/15$:
\begin{eqnarray}
{\rm Tr}  \; G \sim \;  \sigma_1  :e^{i \sqrt{2 \pi} \; {\vec \omega}_1 \cdot {\vec \Phi}}:
+ \sigma_2  :e^{- i \sqrt{2 \pi} \; {\vec \omega}_2 \cdot {\vec \Phi}}:
+ \sigma_3  :e^{i \sqrt{2 \pi} \; \left( - {\vec \omega}_1 + 
{\vec \omega}_2 \right)\cdot {\vec \Phi}}: ,
\label{TraceWZNWSU3}
\end{eqnarray}
where  ${\vec \omega}_{1,2}$ are the fundamental weights of su(3) with the property
${\vec \omega}_{i}^2 = 2/3, {\vec \omega}_{1} \cdot  {\vec \omega}_{2} = 1/3$.
The bosonic field ${\vec \Phi}$ is a compactified field with the following redundancy according to 
Eq. (\ref{gepnerprimary}):
\begin{eqnarray}
{\vec \Phi}  \sim {\vec \Phi} + \sqrt{2 \pi}  \; ( n_1 {\vec \alpha}_1 +   n_2 {\vec \alpha}_2),
\label{identiSu3}
\end{eqnarray}
$n_i$ being integers and the identification (\ref{identiSu3}) involves  the root lattice Q which is
 generated by the simple roots ${\vec \alpha}_{1,2}$ (${\vec \alpha}_{i}^2 = 2$)  of su(3).
One can repeat the analysis with the conjugate representation of SU(3) with highest weight $\vec \Lambda = (0,1)$
to find: 
\begin{eqnarray}
{\rm Tr}  \; G^{\dagger} =  \;  \sigma_1  :e^{- i \sqrt{2 \pi} \; {\vec \omega}_1 \cdot {\vec \Phi}}:
+ \sigma_2  :e^{i \sqrt{2 \pi} \; {\vec \omega}_2 \cdot {\vec \Phi}}:
+ \sigma_3  :e^{i \sqrt{2 \pi} \; \left( {\vec \omega}_1 -
{\vec \omega}_2 \right)\cdot {\vec \Phi}}: .
\label{TraceconjWZNWSU3}
\end{eqnarray}

We now consider the adjoint representation of su(3) with dimension eight. The highest weight for the latter representation
is ${\vec \Lambda} = (1,1)$ with weights $\{(1,1), (-1,2), (2,-1), (0,0), (-2,1), (1, -2), (-1,-1)\}$, the weight (0,0) being doubly degenerate.
Using the decomposition (\ref{gepnerprimary}), model (\ref{model}) for $k=2$ and $N=3$ reads then as follows:
\begin{eqnarray}
{\cal H}_{\rm eff} &=&  {\cal H}^{0}_{SU(3)_2/U(1)^{2}} + \frac{1}{2} \left( \left(\partial_x {\vec \Phi} \right)^2
+   \left(\partial_x {\vec \Theta} \right)^2 \right) \nonumber \\
&+& \lambda \;  \rho +   \lambda  \;  \left( \sigma_1  :e^{i \sqrt{2 \pi} \; {\vec \alpha}_2 \cdot {\vec \Phi}}:
 + \sigma_2  :e^{i \sqrt{2 \pi} \;  {\vec \alpha}_1 \cdot {\vec \Phi}}:
 + \sigma_3  :e^{i \sqrt{2 \pi} \;  {\vec \alpha}_3 \cdot {\vec \Phi}}: + H.c. \right) ,
\label{effbisGPSU3}
\end{eqnarray}
where $ {\cal H}^{0}_{SU(3)_2/U(1)^{2}} $ is the Hamiltonian density of the SU(3)$_2/ U(1)^{2}$ CFT 
and ${\vec \Theta} $ is the dual field of ${\vec \Phi}$.
Expression (\ref{effbisGPSU3}) can also be reproduced by considering  the  OPE
${\rm Tr}  \; G \; {\rm Tr}  \; G^{\dagger} $ from
Eqs. (\ref{TraceWZNWSU3}, \ref{TraceconjWZNWSU3}) and the fusion rules (\ref{GPSU3fusion}).

The IR properties of model (\ref{effbisGPSU3}) are difficult to determine.
In close parallel to the $N=2$ case  (see Appendix B), we single out the perturbation with the $\rho$ field which
takes the form of the so-called homogenous sine-Gordon (HSG) model:
\begin{eqnarray}
{\cal H}_{\rm HSG} =  {\cal H}^{0}_{SU(3)_2/U(1)^{2}} 
+  \lambda \;  \rho .
 \label{HSG}
\end{eqnarray}
The latter is an integrable massive field theory for all sign of $\lambda$ and corresponds to 
the generalization of  the integrable model of ${\mathbb{Z}}_k$ parafermions
perturbed by their first thermal operator \cite{miramontes,HSGbib,dorey}.
When $ \lambda  <0$, i.e. $\langle \rho \rangle >0$, we have in our conventions $\langle \sigma_i \rangle \ne 0$ 
($\sigma_i \times \sigma_i \sim I + \rho$). At lower energy than the mass gap of the HSG model, model (\ref{effbisGPSU3}) becomes equivalent to a double sine-Gordon model with scaling dimension one:
\begin{eqnarray}
{\cal H}_{\rm SG} =   \frac{1}{2} \left( \left(\partial_x {\vec \Phi} \right)^2
+   \left(\partial_x {\vec \Theta} \right)^2 \right)
-    {\tilde \lambda}  \; \sum_{i=1}^{3} \cos \left( \sqrt{2 \pi} \;  {\vec \alpha}_i \cdot {\vec \Phi} \right).
\label{doubleSG}
\end{eqnarray}
The latter model has a spectral gap
and the bosonic ${\vec \Phi}$ field is pinned in the minima of the sine-Gordon potential (${\tilde \lambda}  > 0$):
\begin{eqnarray}
\langle {\vec \Phi}  \rangle =  \sqrt{2 \pi}  \; ( m_1 {\vec \omega}_1 +   m_2 {\vec \omega}_2),
\label{pinninSu3}
\end{eqnarray}
$m_i$ being integers. The vacuum expectations values of the bosons are thus described by the weight lattice $P$
which is generated by the fundamental weights. Using the identification (\ref{identiSu3}) on the bosons,
we find that the ground state is only three-fold degenerate since the ratio P/Q  of the lattices is isomorphic to the center of su(3): $P/Q \sim {\mathbb{Z}}_3$ \cite{dms}.

When  $ \lambda   > 0$, we have now $\langle \rho \rangle <0$.  From the fusion rule (\ref{GPSU3fusion})
$\sigma_i \times \sigma_i \sim I + \rho$, we expect that one enters a phase where 
$\langle \sigma_i \rangle = 0$ and the disorder fields $\mu_i$ of GP should condense now. 
Using the fusion rules (\ref{GPSU3fusion}), we need to consider the second-order in
perturbation theory to generate an effective low-energy model for the ${\vec \Phi}$  bosons
after the integration of the massive GP degrees of freedom:
\begin{eqnarray}
{\cal H}^{\rm SG}_{\rm eff} = \frac{v}{2} \sum_{i=1}^{2} \left( \frac{1}{K_i} \left(\partial_x  \Phi_i  \right)^2
+  K_i  \left(\partial_x \Theta_i \right)^2 \right) 
 - g \;  \sum_{i=1}^{3} \cos \left( \sqrt{8 \pi} \;  {\vec \alpha}_i \cdot {\vec \Phi} \right) ,
\label{SGmodel2ndorder}
\end{eqnarray}
where $g \simeq \lambda^2 >0$ and $K_i$ are the Luttinger parameters for the bosonic fields $\Phi_i$. 
The scaling dimension of the perturbation is $4$ and thus irrelevant. 
One expects the emergence of a quantum critical behavior with two gapless bosons, i.e., 
with central charge $c=2$. 
However, we need to determine the actual values of the Luttinger parameters  which should 
be fixed since the symmetry of the initial model (\ref{effbisGPSU3}) is SU(3).

When $ \lambda >0$, we have $\langle \sigma_i \rangle = 0$ and  ${\rm Tr}  \; G$ in Eq. (\ref{TraceWZNWSU3})
seems to be naively short ranged. However, we need to 
considering higher-order in perturbation theory to deduce the low-energy expression of ${\rm Tr}  \; G$.
In this respect, by performing the OPE between (\ref{TraceWZNWSU3})  and the interacting Hamiltonian (\ref{effbisGPSU3}), we see that the contribution of the $\sigma_i$ fields disappear and we 
get the IR representation of  ${\rm Tr}  \; G$:
\begin{eqnarray}
{\rm Tr}  \; G &\sim&  \;    :e^{- i \sqrt{2 \pi} \; \left( {\vec \omega}_1 - {\vec \alpha}_2 \right) \cdot {\vec \Phi}}:
+  :e^{- i \sqrt{2 \pi} \; \left( {\vec \omega}_2 - {\vec \alpha}_1 \right) \cdot {\vec \Phi}}:
+  :e^{i \sqrt{2 \pi} \; \left( - {\vec \omega}_1 + 
{\vec \omega}_2 - {\vec \alpha}_1 -  {\vec \alpha}_2 \right)\cdot {\vec \Phi}}: 
 \nonumber \\
&+& :e^{i \sqrt{2 \pi} \; \left( {\vec \omega}_1 + {\vec \alpha}_2 \right)\cdot {\vec \Phi}}:
+  :e^{- i \sqrt{2 \pi} \; \left( {\vec \omega}_2 + {\vec \alpha}_1 \right) \cdot {\vec \Phi}}:
+  :e^{i \sqrt{2 \pi} \; \left( - {\vec \omega}_1 + 
{\vec \omega}_2 + {\vec \alpha}_1 + {\vec \alpha}_2 \right)\cdot {\vec \Phi}}: \nonumber \\
 &\sim&  \; 
 :e^{i \sqrt{8 \pi} \;  {\vec \omega}_2 \cdot {\vec \Phi}}: 
 + :e^{- i \sqrt{8 \pi} \;  {\vec \omega}_1 \cdot {\vec \Phi}}: 
+ :e^{i \sqrt{8 \pi} \;  \left( {\vec \omega}_1 - {\vec \omega}_2 \right) \cdot {\vec \Phi}}: .
\label{TraceWZNWSU3IR}
\end{eqnarray}
Since the three terms should have the same scaling dimension, we have necessarily $K_1 = K_2=K$
and the scaling dimension of ${\rm Tr} \; G$ in the IR limit is: $4K/3$.  Since we expect a massless flow
to SU(3)$_1$ in the far IR, the SU(3)$_2$ WZNW primary field G will transmute to the  SU(3)$_1$ one with
scaling dimension $2/3$. We have thus $K_1 = K_2 = 1/2$. 
The double sine-Gordon model (\ref{SGmodel2ndorder}), which describes the IR physics of
model (\ref{model}) when $\lambda >0$,  becomes marginal and identifies 
with the marginal irrelevant SU(3)$_1$ current-current perturbation \cite{andrei}.
In summary, model (\ref{model}) for $k=2$ and $N=3$ 
displays a massless RG flow to SU(3)$_1$ when $\lambda >0$ with the ultraviolet-infrared transmutation:
\begin{eqnarray}
 {\rm Tr}  \Phi_{\rm adj}  &\rightarrow& - J_R^A J_L^A \nonumber \\
 G_{SU(3)_2} &\rightarrow& G_{SU(3)_1} ,
\label{UVIRSU3}
\end{eqnarray}
$J_{R,L}^A$ being the SU(3)$_1$ chiral currents and $G_{SU(3)_1}$ is the SU(3)$_1$ WZNW primary field.

\section{Concluding remarks}

In this work, we have identified an IR massless RG flow for the SU($N$)$_k$ WZNW model perturbed by 
its relevant adjoint primary field. Using a mapping onto $k$-leg 
SU($N$) spin ladder and assuming a weak-strong  coupling continuity, we have shown that this model has critical properties in the SU($N$)$_1$ universality class 
when $N$ and $k$ have no common divisor. This result is the generalization to SU($N$)
of the Haldane's conjecture on spin-$S$ SU(2) Heisenberg chain whose physical properties are governed by 
the SU(2)$_{2S}$ WZNW model perturbed by its adjoint primary field. 
The massless RG flow, presented in this paper, is consistent with the extension of the selection rule on WZNW models based on the global anomaly of the ${\mathbb{Z}}_2$ symmetry \cite{oshikawa}.
Futhermore, we have confirmed the existence of the massless RG
flow for the special $N=3$ and $k=2$ case by means of a direct approach using Gepner's parafermions. 
The resulting non-trivial IR fixed point for $k=2$ can be investigated by a perturbative RG approach in the 
large-$N$ limit since the scaling dimension of the adjoint primary field is close to two when $N \gg 1$.

As perspectives, it will be interesting to have a direct approach of the IR properties of the perturbed SU($N$)$_k$ WZNW 
to complement the analysis of the massless RG flow identified in this paper. In this respect, a truncated conformal space approach in the simplest $N=3$ and $k=2$ case will be very useful as it has been done recently for SU(2)$_k$ perturbed CFT
 \cite{tsvelik,koniksierra}.  A semiclassical approach of model (\ref{model}) 
might also be very helpful to interpret the massless RG flow reported
in this work as the result of some non-linear sigma model with a topological term 
as in the $N=2$ case \cite{affleckhaldane}.
Finally, direct numerical calculations of $k$-leg  SU($N$) spin ladder with $J_{\perp} <0$ and SU($N$) spin chain
models with symmetric rank-$k$ tensor representation might also be fruitful to reveal the massless RG flow.
We hope that future studies will shed light on theses questions and others results on SU($N$)$_k$  perturbed
CFT will be obtained.

\begin{acknowledgements}
The author is grateful to R. Santachiara, A. M. Tsvelik, V. Bois, S. Capponi,  T. Quella, H. Saleur,
K. Totsuka, and A. Weichselbaum  for useful discussions.
\end{acknowledgements}


\appendix

\section{Relation to SU($N$) spin chain problems with symmetric representations}

We relate here the field theory  (\ref{model}) to an SU($N$) spin chain problem
where the spin operators belong in symmetric rank-$k$ tensor representation of SU($N$).
Such mapping might be useful to investigate numerically the massless RG flow proposed
in this paper.

The starting point of the analysis is the existence of an integrable SU($N$) model
with degrees of freedom in symmetric rank-$k$ tensor representation, introduced by Andrei
and Johannesson (AJ) \cite{andreiJ,johannesson}.
The latter model is the SU($N$) generalization of Bethe-ansatz integrable spin-$S$ Heisenberg chain models
which represent unstable SU(2)$_{2S}$ quantum critical points \cite{affleckschulz, affleckhaldane}.
It has been numerically proved that the AJ model displays a quantum critical behavior in
the general SU($N$)$_k$ WZNW universality class \cite{alcaraz,rachel}.
Since the latter CFT has many relevant primary fieds, the critical point is expected to be very fragile.
In this respect, let us introduce the following bilinear-biquadratic SU($N$) 
spin chain model to analyse the stability of the AJ model  in the simplest $k=2$ case:
\begin{eqnarray}
{\cal H} &=& {\cal H}_{\rm AJ} + \delta  \sum_i  \left(S^{A}_{i}  S^{A}_{i+1} \right)^2,
\nonumber \\
 {\cal H}_{\rm AJ}  &=&
   \sum_i  \left( S^{A}_{i}  S^{A}_{i+1} - \frac{N}{3N - 4} \left(S^{A}_{i}  S^{A}_{i+1} \right)^2 \right),
\label{AJmodel}
\end{eqnarray}
where $S^{A}_{i} $ denotes spin operators on site $i$ which transforms in the symmetric 
representation  {\tiny $\yng(2)$} of the SU($N$) group.  Model (\ref{AJmodel}) interpolates between the AJ model for $\delta = 0$ with SU($N$)$_2$ quantum criticality and the pure Heisenberg model  in {\tiny $\yng(2)$} 
representation for $\delta = N/(3N-4)$. As already stressed in section II.C, 
the latter displays an SU($N$)$_1$ quantum critical behavior when $N$ is odd \cite{affleck,greiter}.
Model (\ref{AJmodel}) might thus be a lattice description of a massless RG flow from 
SU($N$)$_2$  to  SU($N$)$_1$ when $N$ is odd. In this respect, it will be interesting to investigate
this model numerically by means of the density-matrix renormalization group calculations in the 
simplest $N=3$ case to shed light on this intriguing possibility.

A field theory analysis of this problem can be obtained by considering at $\delta=0$ an 1D U($N$) fermionic Hubbard model with fermions with $N$ flavors and $k$ colors at filling factor $1/N$ \cite{affleck}.
For this special filling, the latter model enjoys an ${\mathbb{Z}}_N$ symmetry, associated to the one-step
translation invariance,  which might protect it from a mass-gap generation \cite{affleck}.
The low-energy approach of model (\ref{AJmodel}) can then be derived when $\delta \ll 1$ by exploiting
the fact that the AJ model has an SU($N$)$_2$ critical behavior:
\begin{equation}
{\cal H} =  \frac{2\pi v}{N+2}\Big(:I_R^A I_R^A: + :I_L^A I^A_L:\Big) + \delta \;  \Phi,  
\label{AJmodelcont}
\end{equation}
where $I_{R,L}^A$ are SU($N$)$_2$  chiral currents and $\Phi$ is the leading relevant perturbation to be found which describes the departure from the SU($N$)$_2$ fixed point.
It should be invariant under the symmetries of the lattice model (\ref{AJmodel}), in particular,  the one-step translation symmetry which takes the form of an ${\mathbb{Z}}_N$ symmetry. 
The latter gives strong constraints on the identification of $\Phi$.  

In this respect, let us recall
the spectrum of the SU($N$)$_2$  CFT. It has $N(N+1)/2$ primary operators with highest-weights ${\vec \Lambda} = \sum_{i=1}^{N-1} \lambda_i  {\vec \omega}_i$ such that the Dynkin labels satisfy the constraint: $ \sum_{i=1}^{N-1} \lambda_i   \le 2$. Introducing $l_i = \sum_{j=i}^{N-1}  \lambda_j$ as a Young tableau row lengths, we see that the Young tableau cannot have more than two columns. The scaling dimensions of the
SU($N$)$_2$  primary fields are given by \cite{schnitzer}:
\begin{eqnarray}
\Delta_{\lambda} = \frac{X + r (N+1) - r^2/N}{N+2},
\label{scalingdimSUN2}
\end{eqnarray}
with $r =  \sum_{i=1}^{N-1} l_i$ is the number of boxes in the Young tableau 
and $X =   \sum_{i=1}^{N-1}   l_i ( l_i - 2 i)$.
The primary field which transforms according to the $k$th antisymmetric representation of SU($N$) is described
by the highest-weight $(0, \ldots, 1, \ldots, 0)$  ($1$ being in the $k$th position) and has scaling dimension
according to Eq. (\ref{scalingdimSUN2}): $\Delta_k = k (N-k)(N+1)/N(N+2)$. These operators cannot
appear in model (\ref{AJmodelcont}) since they are not invariant under the one-step translation symmetry.
The most relevant operator, which is translational invariant, turns out to be the primary field in the adjoint representation 
with highest weight (1, 0, \ldots, 0, 1) and scaling dimension $2N/(N+2)$. 
We thus expect that the continuum limit of model (\ref{AJmodel}) in the
vicinity of $\delta =0$ is described by the field theory (\ref{model}). The massless flow for odd $N$
from SU($N$)$_2$ to SU($N$)$_1$ can thus be directly investigated numerically  from the 1D lattice model  (\ref{AJmodel})  
by means of density matrix renormalization group calculations for instance.

\section{${\mathbb{Z}}_k$ parafermionic approach to the $N=2$ case}

In this Appendix, we review for completeness the IR properties of model  (\ref{model}) with $N=2$  
by using a mapping onto ${\mathbb{Z}}_k$ parafermionic CFT \cite{cabra}.
In the $N=2$ case,  model  (\ref{model})  reads as follows:
\begin{equation}
{\cal H} = \frac{2\pi}{2+k}\Big(:{\vec I}_R \cdot {\vec I}_R: + :{\vec I}_L \cdot {\vec I}_L:\Big) + \lambda \;  {\rm Tr}  \; \Phi^{(1)} ,
\label{modelN2}
\end{equation}
where ${\vec I}_{R,L}$ are chiral SU(2)$_k$  currents and 
$\Phi^{(1)}$ is the spin-1 SU(2)$_k$ primary field with scaling dimension $4/(k+2)<2$.
We then consider the conformal embedding \cite{para}:
\begin{equation}
SU(2)_k \sim {\mathbb{Z}}_k \times U(1) ,
\label{pararemb}
\end{equation}
where the ${\mathbb{Z}}_k$ parafermionic CFT with central charge  $c= 2 (k-1)/(k+2)$  describes the critical properties of 
two-dimensional ${\mathbb{Z}}_k$ generalization of the Ising model \cite{para}.
The SU(2)$_k$ primaries ($\Phi_{m, \bar m}^{(j)}$)
are related to the Z$_k$ parafermionic ones ($f_{m, \bar m}^{l}$) by \cite{para,gepner}:
\begin{equation}
\Phi_{m, \bar m}^{(j)} = f_{2m,2 \bar m}^{l} :\exp\left(i \; m \sqrt{\frac{8\pi}{k}}
\; \Phi_L+  i \; \bar m \sqrt{\frac{8\pi}{k}} \; \Phi_R\right):,
\label{suparaprim}
\end{equation}
where $l=2 j = 0,..,k$ and $\Phi$ is a Bose field with chiral components $\Phi_{R,L}$.
From the identification (\ref{suparaprim}), we get:
\begin{eqnarray}
{\rm Tr}  \; G &=& {\rm Tr}  \; \Phi^{(1/2)} \sim \sigma_1  :e^{i \;  \sqrt{\frac{2\pi}{k}} \Phi}: +
\ \sigma_1^{\dagger}  :e^{-i \;  \sqrt{\frac{2\pi}{k}} \Phi}:
\nonumber \\
{\rm Tr}  \; \Phi^{(1)} &\sim& \epsilon_1 + 
\sigma_2  :e^{i \;  \sqrt{\frac{8\pi}{k}} \Phi}: +
\ \sigma_2^{\dagger}  :e^{-i \;  \sqrt{\frac{8\pi}{k}} \Phi}: ,
\label{paraboso}
\end{eqnarray}
where $\sigma_p$  ($\sigma_p^{\dagger} = \sigma_{k-p}$) are ${\mathbb{Z}}_k$
primary fields with scaling dimensions $d_p = p(k-p)/k(k+2)$ $(p=0, \ldots, k -1)$ 
which describe long-distance correlations of the ${\mathbb{Z}}_k$ Ising lattice spins.
In Eq. (\ref{paraboso}), $ \epsilon_1$ is the first thermal operator with scaling dimension $4/(k+2)$
and our notations are such that  $\langle  \epsilon_1 \rangle > 0 $
in the ${\mathbb{Z}}_k$ ordered phase of the underlying ${\mathbb{Z}}_k$ Ising model.
Model (\ref{modelN2}) can then be rewritten as follows in this new basis:
\begin{equation}
{\cal H} = {\cal H}^{0}_{{\mathbb Z}_k}   +  \frac{1}{2}  \left( \left(\partial_x  \Phi  \right)^2
+   \left(\partial_x \Theta \right)^2 \right)  +  \lambda   \epsilon_1 + 
 \lambda  \sigma_2  :e^{i \;  \sqrt{\frac{8\pi}{k}} \Phi}: +
 \lambda  \sigma_2^{\dagger}  :e^{-i \;  \sqrt{\frac{8\pi}{k}} \Phi}: ,
\label{modelN2para}
\end{equation}
$\Theta$ being the dual Bose field ($\Theta = \Phi_{L} - \Phi_{R}$). 
The perturbation is strongly relevant and since the ${\mathbb{Z}}_k$ degrees
of freedom are discrete, one expects the opening of a mass gap in this sector.

The next step of the approach is therefore to single out the thermal perturbation on the ${\mathbb{Z}}_k$ 
parafermions: 
 \begin{eqnarray}
{\cal H}_{{\mathbb{Z}}_k} =  {\cal H}^{0} ({\mathbb{Z}}_k)  
+  \lambda \;   \epsilon_1 ,
 \label{fateevmodel}
\end{eqnarray}
which is an integrable massive field theory for all sign of $\lambda$ \cite{fateev}.
When $\lambda <0$, we have $\langle  \epsilon_1 \rangle >0$ and the order spin fields condense $\langle \sigma_p\rangle \ne 0$,  leading thus to a sine-Gordon model for the $\Phi$ field:
\begin{equation}
{\cal H}_{\rm SG} = \frac{1}{2}  \left( \left(\partial_x  \Phi  \right)^2
+   \left(\partial_x \Theta \right)^2 \right)   +
 {\tilde \lambda} \cos \left(\sqrt{\frac{8\pi}{k}} \Phi \right) ,
\label{modelSGSU2lambdapositive}
\end{equation}
which is a massive field theory for all $k$.
The situation turns out to be very
different when  $\lambda >0$. In the latter case, we have now $\langle  \epsilon_1 \rangle < 0$ and the parafermionic belongs to the disordered phase where $\langle \sigma_p\rangle = 0$. 
The integration over the massive parafermionic degrees of freedom in Eq. (\ref{modelN2para})
leads to an low-energy effective Hamiltonian for the bosonic field $\Phi$ which depends on the 
parity of $k$ \cite{cabra,nonne2011}.

\subsection{$k$ odd case}

We first consider the $k$ odd case. The $\sigma_2$ operator in Eq. (\ref{modelN2para}) carries a $p=2$ charge
under  ${\mathbb{Z}}_k$. We need to consider higher-order in perturbation theory to cancel out the ${\mathbb{Z}}_k$  
charge of $\sigma_2$.  When $k$ is odd, the $k$th order of perturbation is necessary 
to suppress the $\sigma_2$ contribution. The low-energy Hamiltonian
for the Bose field $\Phi$ takes then the form of a sine-Gordon model at $\beta^2 = 8 \pi k$:
\begin{eqnarray}
\mathcal{H}^{\rm odd}_{\rm eff} = \frac{v}{2} \left( \frac{1}{K} \left(\partial_x \Phi\right)^{2}
+ K \left(\partial_x \Theta\right)^{2} \right) 
+ g  \cos \left(\sqrt{8 \pi k} \; \Phi \right),
\label{hcNodd}
\end{eqnarray}
with $g  \sim \lambda^k$, $K$ is the Luttinger parameter and $v$ is a velocity.  The scaling dimension of the perturbation
is $2k K$ and thus naively irrelevant. We also notice that the approach gives the same low-energy
effective field theory  than the one derived directly 
by Schulz from the Abelian bosonization of half-integer $S= k/2$  Heisenberg spin chain \cite{schulz}.
The global continuous symmetry of model (\ref{modelN2para})
is SU(2) and therefore the Luttinger parameter $K$ should be fixed to a value compatible with this non-Abelian symmetry.
One way to identify $K$ is to determine the IR limit of ${\rm Tr}  \; G $ in Eq. (\ref{paraboso}). 
By fusing this operator with the Hamiltonian (\ref{modelN2para}) at the $(k-1)/2$th order in perturbation theory,
the $\sigma_1$ contribution disappears and one obtains the low-energy limit description of  ${\rm Tr}  \; G $:
\begin{equation}
{\rm Tr}  \; G  \sim  \cos( \sqrt{2\pi k K} \Phi) ,
\label{GN2IRlimt}
\end{equation}
which corresponds to the SU(2) spin-singlet dimerization operator if $K=1/k$ \cite{bookboso}.
The SU(2) symmetry of the problem fixes thus $K =1/k$ and $G$ has scaling dimension $1/2$, i.e., 
corresponds in the far IR  limit to the spin-1/2 SU(2)$_1$ WZNW primary.
A massless RG flow to SU(2)$_1$ CFT is therefore expected in the odd $k$ case when $\lambda >0$ as
it should be.

\subsection{$k$ even case}

As in the $k$ odd case,  one has to consider higher orders in perturbation theory
to derive an effective theory for the field $\Phi$  since the
$\sigma_2$ operator in Eq.  (\ref{modelN2para}) average to zero in the ${\mathbb{Z}}_k$  symmetric phase.
When $k$ is even, one needs now the $k/2$ th order
of perturbation theory to cancel out the $\sigma_2$ operator in Eq. (\ref{modelN2para}).
The resulting low-energy Hamiltonian then reads as follows:
\begin{eqnarray}
\mathcal{H}^{\rm even}_{\rm eff} = \frac{v}{2} \left( \frac{1}{K} \left(\partial_x \Phi \right)^{2}
+ K \left(\partial_x \Theta \right)^{2} \right) 
+ g  \cos \left(\sqrt{2 \pi k} \; \Phi \right) .
\label{hcNeven}
\end{eqnarray}
We recover the same low-energy approach than the one derived by Schulz in his study of integer Heisenberg
spin $S=k$ chain \cite{schulz}.
As seen in the previous case, the SU(2) symmetry is fixed by the value of the Luttinger parameter: $K=1/k$.
Model (\ref{hcNeven}) becomes the $\beta^2 = 2 \pi$ sine-Gordon model which is massive 
and enjoys an hidden SU(2) symmetry \cite{affleckSU2}.
In stark contrast to the odd $k$ case, one cannot suppress the contribution 
of the spin field $\sigma_1$  in Eq. (\ref{paraboso}) by considering higher-order in perturbation theory with 
the Hamiltonian (\ref{modelN2para}). The IR limit of ${\rm Tr}  \; G $ always gives a short-ranged contribution. 
In summary, from this ${\mathbb{Z}}_k$ parafermionic approach,  we conclude 
 that model (\ref{modelN2}) is fully gapped for all sign of $\lambda$ when $k$ is even as it should.


\end{document}